\pdfoutput=1
\newcommand{\diffunit}{$\mathrm{GeV\;cm^{-2}\;s^{-1}\;sr^{-1}}$}

\newcommand{\Nch}{$N_{\mathrm{ch}}\;$}

\newcommand{\esqdnde}{$\mathrm{E^{2}_{\nu} \times dN_{\nu}/dE_{\nu}}$}

\documentclass[a4paper,superscriptaddress]{jpconf}
\usepackage{graphicx}

\begin{document}
\title{\Large 
Contributions to The $10^{th}$ International Conference on Topics in Astroparticle and Underground Physics (TAUP) 2007, Sendai, Japan, Sep. 11-15, 2007}

\begin{center}
{\bf \large The IceCube Collaboration}
\end{center}

\noindent
R.~Abbasi$^{20}$,
M.~Ackermann$^{32}$,
J.~Adams$^{11}$,
J.~Ahrens$^{21}$,
K.~Andeen$^{20}$,
J.~Auffenberg$^{31}$,
X.~Bai$^{23}$,
M.~Baker$^{20}$,
B.~Baret$^{9}$,
S.~W.~Barwick$^{16}$
% ...
\\

\noindent
% ...
$^{9}$ Vrije Universiteit Brussel, Dienst ELEM, B-1050 Brussels, Belgium \\
% ...
$^{11}$ Dept.~of Physics and Astronomy, University of Canterbury, Private Bag 4800, Christchurch, New Zealand \\
% ...
$^{16}$ Dept.~of Physics and Astronomy, University of California, Irvine, CA 92697, USA \\
% ...
$^{20}$ Dept.~of Physics, University of Wisconsin, Madison, WI 53706, USA \\
$^{21}$ Institute of Physics, University of Mainz, Staudinger Weg 7, D-55099 Mainz, Germany \\
% ...
$^{23}$ Bartol Research Institute and Department of Physics and Astronomy, University of Delaware, Newark, DE 19716, USA \\
% ...
$^{31}$ Dept.~of Physics, University of Wuppertal, D-42119 Wuppertal, Germany \\
$^{32}$ DESY, D-15735 Zeuthen, Germany \\
% ...

\newpage
{\bf Acknowledgments} \\

We acknowledge the support from the following agencies:
National Science Foundation-Office of Polar Program,
National Science Foundation-Physics Division,
University of Wisconsin Alumni Research Foundation,
Department of Energy, and National Energy Research Scientific Computing Center
(supported by the Office of Energy Research of the Department of Energy),
the NSF-supported TeraGrid system at the San Diego Supercomputer Center (SDSC),
and the National Center for Supercomputing Applications (NCSA);
Swedish Research Council,
Swedish Polar Research Secretariat,
and Knut and Alice Wallenberg Foundation, Sweden;
German Ministry for Education and Research,
Deutsche Forschungsgemeinschaft (DFG), Germany;
Fund for Scientific Research (FNRS-FWO),
Flanders Institute to encourage scientific and technological research in industry (IWT),
Belgian Federal Office for Scientific, Technical and Cultural affairs (OSTC);
the Netherlands Organisation for Scientific Research (NWO);
M.~Ribordy acknowledges the support of the SNF (Switzerland);
A.~Kappes and J.~D.~Zornoza acknowledge support by the EU Marie Curie OIF Program.
\newpage
\begin{center}
{\bf \Large Table of Contents}
\end{center}
\vskip 2 cm
\begin{itemize}
\item[\bf 1.] K.~Hoshina for the IceCube Collaboration, {\it Diffuse high-energy neutrino searches in AMANDA-II and IceCube: Results and future prospects}
\item[\bf 2.] P.~Desiati for the IceCube Collaboration, {\it Measurement of the atmospheric neutrino flux with AMANDA-II and IceCube}
\item[\bf 3.] T.~Montaruli for the IceCube Collaboration, {\it First Results of the IceCube Observatory on High Energy Neutrino Astronomy}
\item[\bf 4.] C.~Rott for the IceCube Collaboration, {\it Indirect Searches for Dark Matter with IceCube}
\item[\bf 5.] M.~Ono, A.~Ishihara and S.~Yoshida for the IceCube Collaboration, {\it Identification of Extremely-high energy starting neutrino events with the IceCube observatory}
\end{itemize}
\newpage
\title{Diffuse high-energy neutrino searches in AMANDA-II and IceCube: Results and future prospects}

\author{Kotoyo Hoshina for the IceCube Collaboration\footnote{IceCube web site at http://icecube.wisc.edu}}

\address{Department of Physics, University of Wisconsin, Madison, U.S.A.}

\ead{hoshina@icecube.wisc.edu}

\begin{abstract}
The AMANDA-II data collected during the period 2000--2003 have been analysed in a search
for a diffuse flux of high-energy extra-terrestrial muon neutrinos from
the sum of all sources in the Universe. With no excess events seen, an
upper limit of \esqdnde $ <7.4\times10^{-8}$ \diffunit was obtained.
The sensitivity of the diffuse analysis of IceCube 9 string for 137 days of data is calculated 
to be \esqdnde $ <1.3\times10^{-7}$ \diffunit. No excess events are observed, which confirms the
AMANDA-II upper limit.

\end{abstract}

\section{Introduction}
Extra-terrestrial neutrinos have been regarded as one of the most promising 
tools to investigate the non-thermal Universe. 
Due to their low interaction cross section, neutrinos retain 
their original source information such as direction and energy 
spectrum.
However, the low interaction cross section makes it difficult to detect neutrinos
which must interact within or near the telescope.
The size of the telescope must thus be large enough to collect a sufficient
number of neutrinos to find and probe extra-terrestrial sources.

AMANDA-II is a neutrino telescope consisting of
677 optical modules attached to 19 strings, deployed
between 1500 m to 1950 m depth within a 200 m diameter in glacial ice at the South Pole.
Following the success of AMANDA-II, the cubic kilometer-sized IceCube experiment
started construction in 2005.
By the spring of 2007, 22 strings with 1320 digital optical modules (DOMs) 
out of the 4800 planned had been deployed, at a depth of 1450 m - 2450 m
surrounding AMANDA-II.

With these observatories numerous analyses are performed to search for 
extra-terrestrial neutrinos~\cite{TAUPteresa,TAUPcarsten,TAUPmio}.
The target of diffuse analysis are neutrinos from unresolved sources.
If the neutrino flux from an individual source is too
small to be detected by point source search techniques~\cite{TAUPteresa}, it is nevertheless
possible to investigate their characteristics by combining events from isotropically distributed
sources.
The extra-terrestrial neutrinos are predicted to follow a \mbox{$\Phi
\propto $ E$^{-2}$} energy spectrum resulting from shock acceleration
processes~\cite{wb_bound}. 
Since the atmospheric neutrino flux has a much softer energy spectrum, 
an excess of events at higher energy over the expected atmospheric neutrino background
would be indicative of an extra-terrestrial $\nu$ flux.

In this paper we present analyses with AMANDA-II data collected during the 
period 2000--2003 and IceCube data obtained with 9 strings (IC9) in 2006. 

\section{Search Method}
Cosmic ray interactions in the atmosphere create pions, kaons, and charmed
hadrons which can later decay into muons and neutrinos. The main background
for this analysis consists of atmospheric muons traveling downward through
the ice. Diffuse analyses use the Earth as a filter to search for upgoing
astrophysical neutrino-induced events. Once the background muons have been
rejected, the data set mainly consists of neutrino-induced upward
events~\cite{TAUPpaolo}.
To separate atmospheric neutrinos from extra-terrestrial neutrinos,
we make use of the number of optical modules that reported at
least one Cherenkov photon during an event (\Nch$\!$) as the energy estimator.
The number of events above an \Nch cut has been compared with 
sets of monte-carlo prediction for E$^{-2}$ and atmospheric neutrino models
(Bartol~\cite{bartol2004,bartol2006}, Honda~\cite{honda2004}). 
The \Nch cut was optimized to produce the best limit 
setting sensitivity~\cite{mrp}.
In order not to bias the analysis, data above the resulting \Nch cut were
kept hidden from analyzer while the lower \Nch
events were compared to atmospheric neutrino expectations from Bartol
and Honda.

\section{AMANDA-II diffuse muon searches}

\begin{figure}
\begin{minipage}{18pc}
\includegraphics[width=17.5pc]{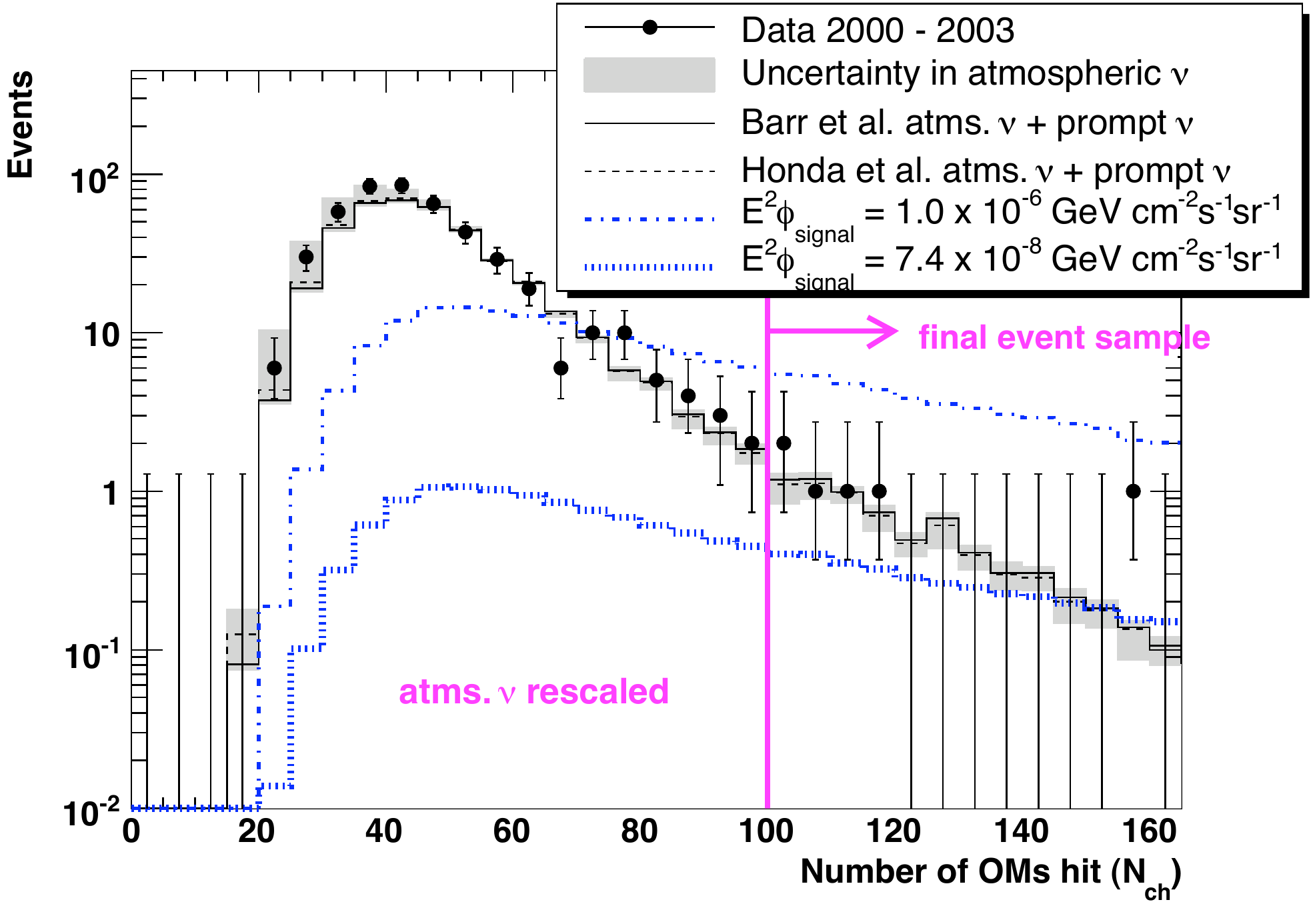}
\caption{
\Nch for the AMANDA-II 2000--2003 diffuse muon
neutrino analysis compared to atmospheric neutrino expectations~\cite{bartol2004,honda2004}. 
}\label{nch_sigrescaled}
\end{minipage}\hspace{2pc}%
\begin{minipage}{18pc}
\includegraphics [width=17.5pc]{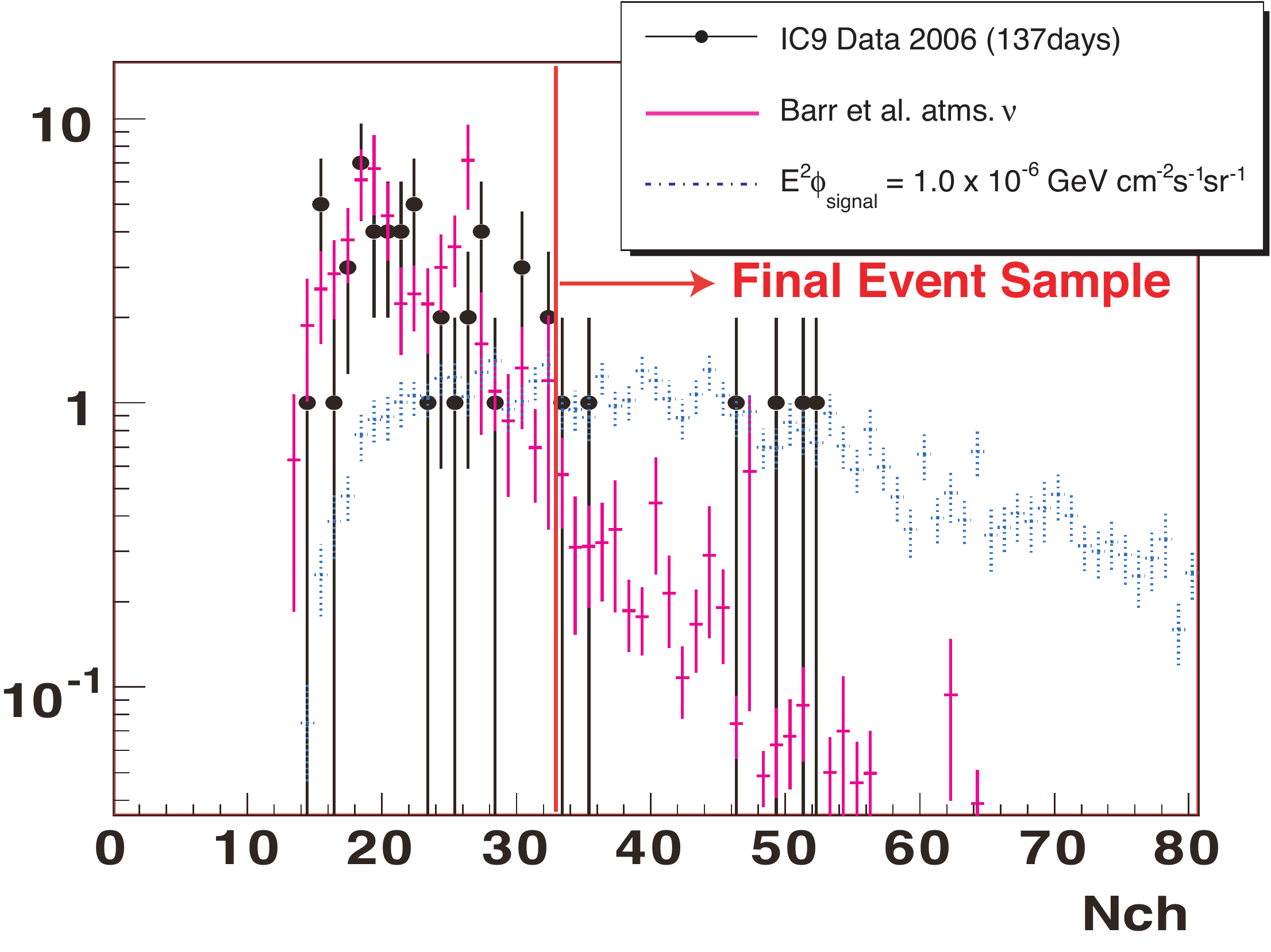}
\vskip -0.5cm
\caption{\Nch distribution in IC9 for 137 days livetime compared to atmospheric neutrino expectations~\cite{bartol2006}.
}\label{ic9nch}
\end{minipage} 
\end{figure}

\begin{figure*}[th!]
\begin{center}
\noindent
\includegraphics [width=0.67\textwidth]{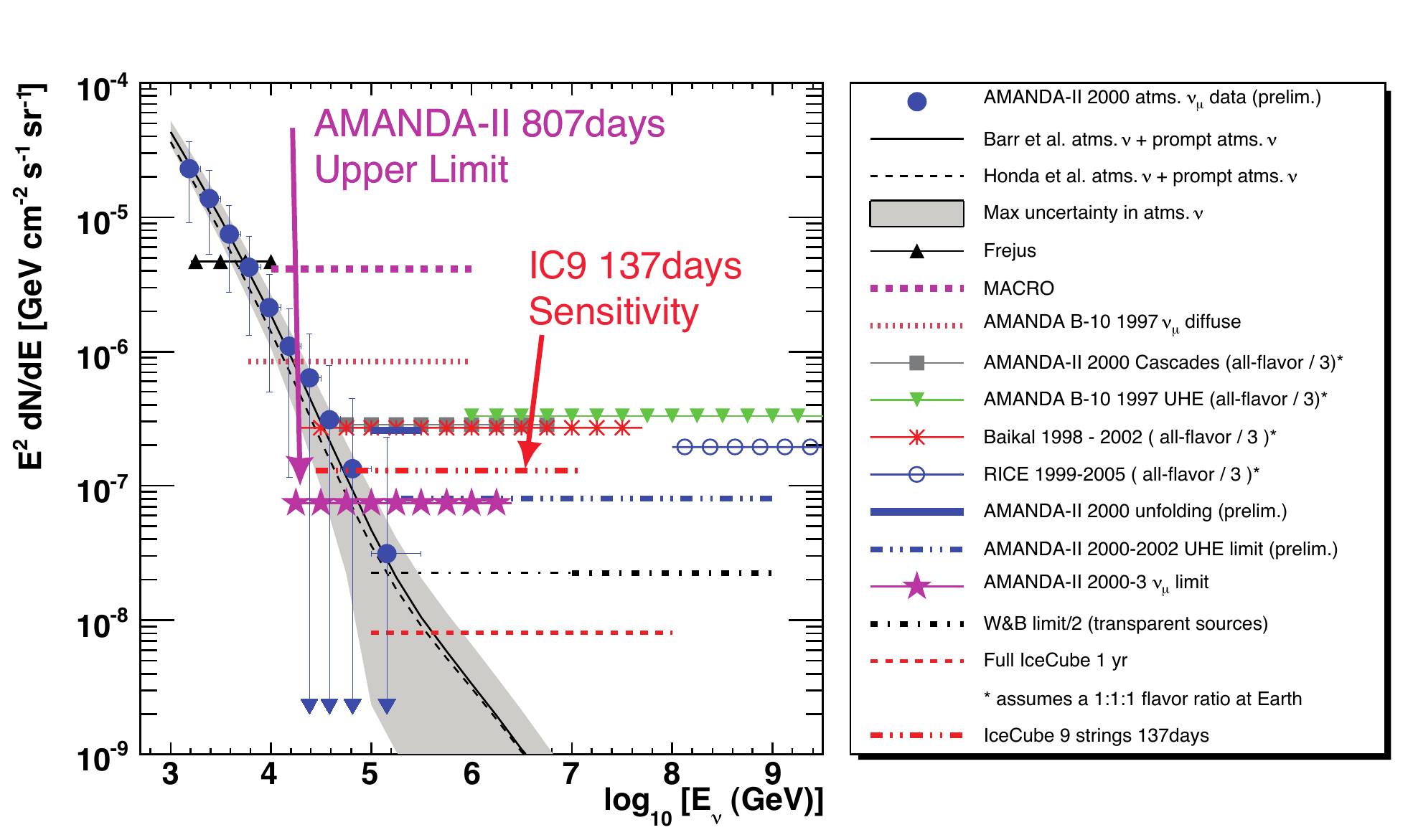}
\end{center}
\vskip -0.5cm
\caption{Upper limit on the $\nu_{\mu}$ flux from sources with an $E^{-2}$ 
energy spectrum are shown for single and all-flavor analyses~\cite{hodges-diffuse,TAUPkirsten}.
}\label{sensitivity}
\end{figure*}

Searches for a diffuse flux have been performed with AMANDA-II data 
obtained 2000-2003 (807 days livetime)~\cite{hodges-diffuse}.
The observed \Nch distribution is compared to the atmospheric neutrino
background calculations in Figure~\ref{nch_sigrescaled}.
For \Nch$>100$ region, 6 events were seen, while 7.0 were
expected. Using the range of atmospheric uncertainty (shaded band in Figure
\ref{nch_sigrescaled}) in the limit calculation~\cite{ch} leads to an upper
limit on a $\Phi \propto E^{-2}$ flux of muon neutrinos at Earth of
\esqdnde $= 7.4\times10^{-8}$ \diffunit. This upper limit is valid in the energy
range 16--2500 TeV. 
Figure~\ref{sensitivity} shows the upper limit 
on the $\nu_{\mu}$ flux from sources with an E$^{-2}$ energy spectrum. 
The limit from the AMANDA-II 807 days analysis is a factor of four above the 
Waxman-Bahcall upper bound~\cite{wb_bound}. 

\vskip -.5cm

\section{IceCube 9 String diffuse muon searches}

Data from the first nine IceCube strings (IC9), with livetime 137 days, was analyzed 
to search for a diffuse flux.
The atmospheric muon rejection and \Nch cut threshold are
re-optimized~\cite{icrc2007} to accommodate the new detector geometry.
The finalized \Nch cut=33 for the IC9 137 days gives a sensitivity
\esqdnde $= 1.3\times10^{-7}$ \diffunit ~on the neutrino energy range of 25 TeV to 10 PeV, 
which is a factor of two above the AMANDA-II 807 day upper limit.

Since the sensitivity of IC9 137 days was well above the AMANDA-II upper limit, 
we revealed the high-energy events (\Nch$>$33) for the sake of a verification study of the detector.
6 events were observed while 6.5 were expected from the Bartol atmospheric neutrino model~\cite{bartol2006}, 
which indicates no signal observation, consistent with the better AMANDA-II upper limit.

\section{Outlook for IceCube 22 strings}

It is an encouraging sign that the IC9 137 day is within a factor of two
of the 807 day AMANDA-II result.
With same atmospheric muon rejection and \Nch cut as IC9, 
a sensitivity of IceCube 22 strings (IC22) for 137 days is expected to reach  
\esqdnde $= 4.0\times10^{-8}$ \diffunit.  
This is already a factor of three better than IC9.
Improvements of event separation to extract extra-terrestrial neutrinos 
from atmospheric neutrinos~\cite{ghill} are ongoing together with 
the optimization of background muon rejection for the IC22 geometry, which should
improve the sensitivity further.

\section{Acknowledgements}
This work is supported by the Office of Polar Programs of the National Science Foundation.

\vskip -2.cm
\title{Measurement of the atmospheric neutrino flux with AMANDA-II and IceCube}

\author{Paolo Desiati$^*$ for the IceCube Collaboration}

\address{$^*$Department of Physics, University of Wisconsin-Madison, 53706, WI, US}

\ead{paolo.desiati@icecube.wisc.edu}

\begin{abstract}
The IceCube Observatory presently consists of an array of 22 vertical strings, deployed between 1450 m and 2450 m of depth and containing 1320 digital optical sensors, and 26 IceTop surface stations with 104 sensors. The denser AMANDA-II array is integrated into IceCube in order to extend the sensitivity to lower energies.  The Observatory is the world's largest neutrino telescope in operation. Recent results on the measurement of the atmospheric neutrino flux with IceCube are discussed, with an emphasis on the acceptance and sensitivity potential of the growing IceCube neutrino telescope.
\end{abstract}

\section{Introduction}
\label{sec:intro}

Neutrinos are the ideal cosmic messengers that allow us to probe the origin of cosmic rays and high energy mechanisms that shaped our Universe, since they can propagate through huge distances almost undisturbed from their sources. Detection of neutrinos in correlation with possible sources of cosmic rays would be the awaited smoking gun that would prove hadronic acceleration. On the other hand, the very same reasons that make neutrinos interesting cosmic messengers, make them also difficult to detect, requiring large volume detectors deployed in the deep oceans or in the antarctic ice \cite{km3}.

The IceCube Observatory is a kilometer-cube size neutrino telescope in construction at the geographic South Pole \cite{icecube}, and it presently consists of 22 strings (hence tagged as IC22). The strings are located on an hexagonal array with a spacing of about 125 m, each of which contains 60 digital optical sensors that are 17 m apart. IceCube embeds the predecessor AMANDA-II array, which consists of 677 analog optical sensors compactly arranged in 19 strings \cite{amanda}.

The goal of neutrino telescopes is the detection of high energy neutrinos of extra-terrestrial origin \cite{here}. The major background is the intense flux of downward muons produced by the impact of cosmic rays on the Earth's atmosphere: about a million of them for each neutrino-induced event.
The virtually background free upgoing atmospheric neutrinos constitute a foreground of diffuse events among which we search for the high energy extra-terrestrial events. The IceCube Observatory will collect an unprecedented statistics of atmospheric neutrinos, allowing us to measure this foreground and to understand it, and, therefore, to do neutrino astrophysics.

\section{Selecting Atmospheric Neutrinos}
\label{sec:atmnu}

\begin{figure}[h]
\begin{minipage}{20pc}
\includegraphics[width=20pc]{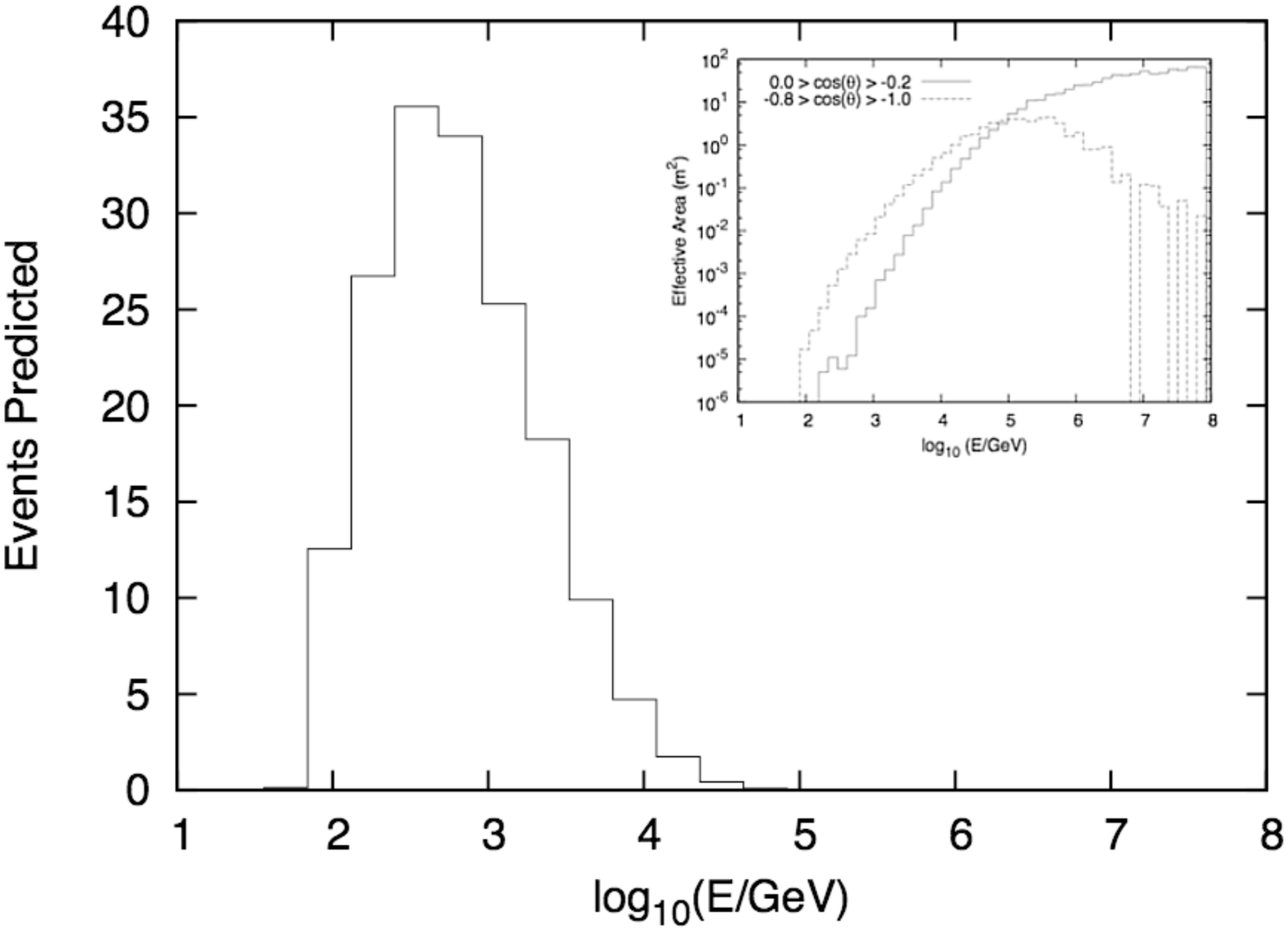}
\vspace{-2pc}
\caption{\label{fig:ic9}Neutrino energy distribution for the event selected in 2006 with 9 strings of IceCube. Insert: the neutrino effective area as a function of neutrino energy.}
\end{minipage}
\hspace{2pc}%
\begin{minipage}{16pc}
\vspace{-2pc}
\includegraphics[width=16pc]{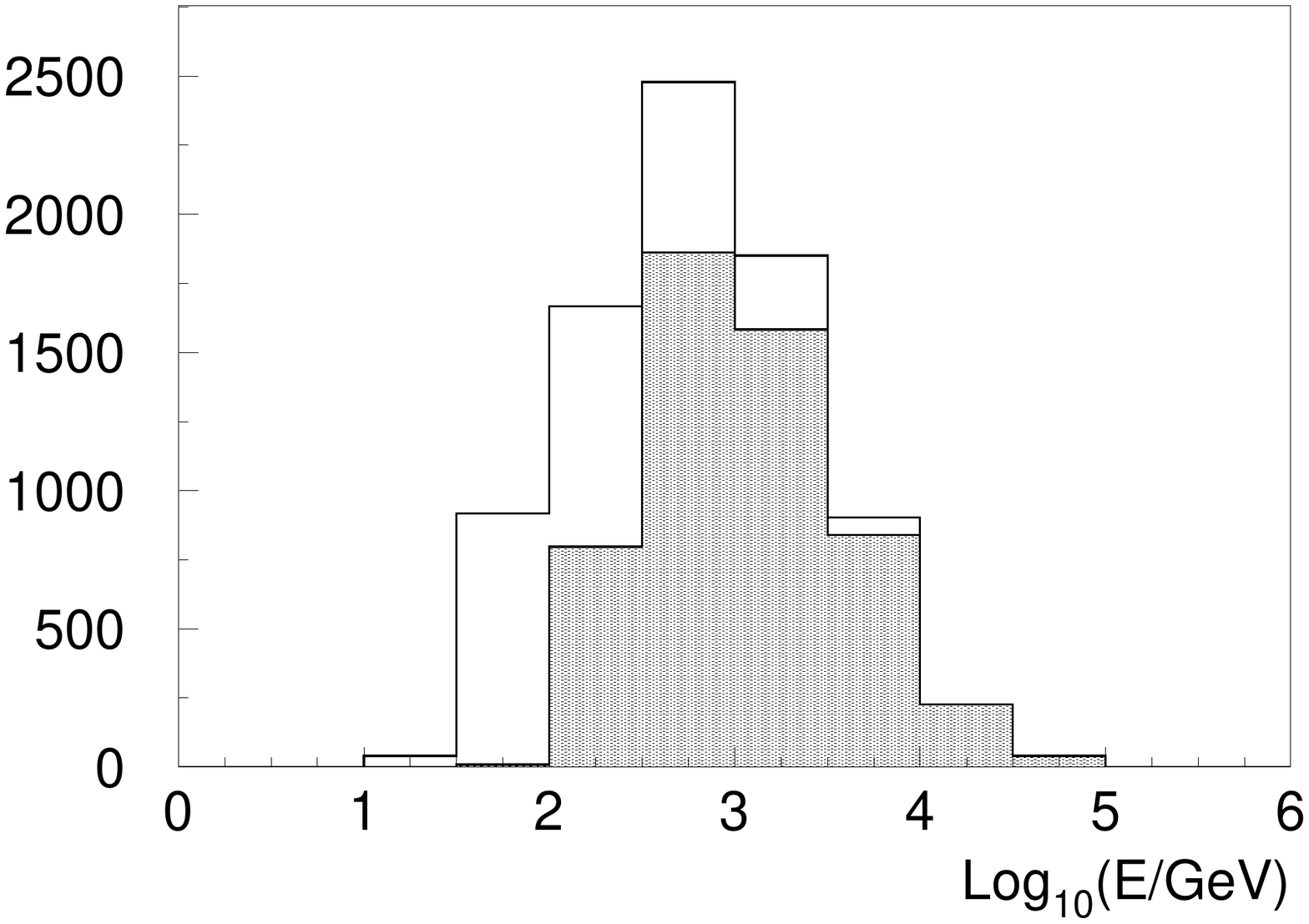}
\vspace{-1pc}
\caption{\label{fig:low}Neutrino energy distribution expected in 200 days with IC22 (shaded area), and with the combined IC22 and AMANDA. In the latter case the energy threshold is about 30 GeV.}
\end{minipage} 
\end{figure}

An upward moving event in a neutrino telescope would be guaranteed to be neutrino-induced muon track, if the direction reconstruction was not affected by any ambiguity. In reality, less than 5\% of the downward muon bundles are mistakenly reconstructed as upgoing. Moreover, in the IC22 array, about 5\% of the recorded events are pairs of uncorrelated muon bundles that happen to pass through the detector in time coincidence. Around 10\% of the coincident events mimic upward moving tracks. Both these sources of background are in excess of about $10^3-10^4$ with respect to the neutrino-induced events, if no specific selection criteria are applied. The event selection is designed to achieve a good track directional resolution. For instance, requiring an angular resolution of about 2$^{\circ}$, we can collect about one atmospheric neutrino per hour in the IC22 array (with a selection efficiency of about 15-20\%), of which about 10-15\% are fakes. The first analysis of atmospheric neutrinos in IceCube was performed on data collected by the first 9 strings during 2006. 234 events were recorded in 137.4 days livetime, with an expectation of 211$\pm$76.1(syst)$\pm$14.5(stat) from Monte Carlo simulations \cite{ic9}. As shown in Fig. \ref{fig:ic9}, most of the selected neutrino events have energy just below 1 TeV. The low energy profile is dominated by the experimental trigger and selection efficiency, while the high energy shape is determined by the steep spectrum and neutrino cross section.

\vspace{-2pc}

The full IceCube array will be able to collect about 5-10 atmospheric neutrinos per hour, depending on the selection criteria and, as a consequence, on the event quality.

The detection of electron neutrino-induced events is more critical than muon events, since their rate is about one order of magnitude lower that the one of muon neutrinos. Such events in IceCube are identified through the hadronic cascades produced by charged current interactions. The main background is due to bremsstrahlung energy loss by muon bundles. A detailed study on background rejection is under way.

\subsection{Low Energy Neutrinos}
\label{ssec:low}

The criteria currently under investigation to achieve a good background rejection, result in a neutrino energy threshold of 100 GeV. At these energies, the uncertainty on the flux of atmospheric neutrinos is estimated to be larger than 30\% \cite{uncertain}. Almost one third of  the uncertainty is due to the composition of cosmic rays above 1 TeV, and the reminder to the high energy hadronic interaction models. Due to kinematical reasons, neutrinos produced in the decays of kaons (K) dominate the atmospheric flux above about 100 GeV \cite{gaisser}, with uncertainties significantly larger than on pion ($\pi$) production in the same energy region. Therefore the measurement of atmospheric neutrinos below 100 GeV would probe a region affected by smaller theoretical uncertainties and would help in constraining the absolute normalization. With theoretical uncertainties below 25\%, we can test the experimental capabilities of the IceCube experiment and assess its systematics down to that order of magnitude.

The IC22 and AMANDA-II arrays can detect events in time coincidence, lowering the neutrino energy threshold to about 30 GeV (resulting in about 50\% more events, see Fig \ref{fig:low}), since AMANDA-II is a denser detector and, combined with IceCube, improves the track angular resolution below 100 GeV. At 30 GeV standard oscillations inevitably affect the measurement of neutrino flux, and for vertical tracks, about 70\% of muon neutrinos have oscillated into another flavor \cite{oscil}. On the other hand, experimental energy and angular resolutions dilute the effect of oscillations. A less systematics prone reconstruction of high quality low energy upward moving tracks close to the IceCube strings, is under way. The muon neutrino survival probability has its minimum at 30 GeV for vertical tracks. Since the optical sensors are only 17 m apart, it is possible to statistically estimate the muon track length $L_{\mu}$ and its energy $E_{\mu}$, and deconvolve them in a L$_{\nu}$/E$_{\nu}$ measurement with minimized systematics due to the $\nu_{\mu}-\mu$ kinematics, where $L_\nu$ is the neutrino path length and $E_\nu$ the neutrino energy.

\subsection{High Energy Neutrinos}
\label{ssec:high}

Due to the steep atmospheric neutrino spectrum most of the events are around a few TeV. Above this energy the detailed measurement of the spectrum slope depends on the primary cosmic ray spectrum and composition, on the K production in the atmosphere and, mainly above 100 TeV and on the yet unknown contribution of charm production. The atmospheric muon flux is also used to probe interaction models. The $\mu^+/\mu^-$ ratio measured by the MINOS experiment at about 1 TeV has provided an important constraint on the K$^+$/K$^-$ ratio in the atmosphere \cite{muratio}. Neutrinos are even more heavily affected by the K ratio above 1 TeV, therefore this constraint is more relevant for neutrino production. With IceCube it will be possible to probe the K/$\pi$ relative contribution to the atmospheric neutrino flux, with a combined analysis of energy and zenith angular shapes.

Atmospheric neutrinos above 10 TeV can be used to probe the internal structure of the Earth, that starts to be opaque to neutrino propagation at about 25 TeV, where the Earth's diameter equals the neutrino absorption length \cite{tomo}. Due to the steepness of the atmospheric neutrino energy spectrum, IceCube will take 10 years to become sensitive to the Earth's matter profile, providing direct geophysical information.
Above 100 TeV neutrinos starts to be dominated by the decay of heavy mesons and baryons containing the charm quarks \cite{henu}. This component of the atmospheric neutrinos is still very uncertain, mainly because of the unknown baryon contribution. A better assessment of the charm background is critical for the discovery of a high energy extra-terrestrial signal.

\section*{References}

\title{First Results of the IceCube Observatory on High Energy Neutrino Astronomy}
\author{Teresa Montaruli$^*$ for the IceCube Collaboration }

\address{$^*$ Department of Physics, University of Wisconsin-Madison, 53706, WI, USA, on leave University of Bari, 70126, Italy}

\ead{teresa.montaruli@icecube.wisc.edu}

\begin{abstract}
The first results of a search for high energy neutrinos in excess of the atmospheric neutrino background, both from any sky direction and from 26 known sources, using 9 strings of IceCube
are presented. We summarize also the results of AMANDA-II from 2000 to 2006. The expected performance of future configurations of the growing IceCube detector are discussed. 
\end{abstract}

\section{Introduction}
The main objective for neutrino telescopes is the detection of astrophysical neutrinos at energies above some tens of GeV, which would prove proton or nuclei acceleration in sources.
This discovery would help in identifying the sources of the ultra-high energy cosmic rays observed up to about $10^{21}$ eV in giant air shower arrays. 
The
analysis techniques illustrated here aim at finding the signal of neutrinos from point-sources on top of the background of atmospheric neutrinos, which is more uniformly distributed in arrival direction.  
We refer to other contributions at this conference for a general introduction to
IceCube and for the detection of diffuse and atmospheric neutrino fluxes \cite{others}.
IceCube is now running with 22 strings (IC22) and during this construction season, just started in Nov. 2007, we aim at installing 14 to 18 more strings. When completed in 2011, it will be composed of up to 80 strings (IC80) holding 4800 photomultipliers at depths between about 1.5 and 2.5 km in the Antarctic ice. 
IceCube can measure all neutrino flavors but muon neutrinos are the only ones that allow a pointing accuracy at the level of a degree.  Hence we focus here on this channel.
Searches for steady-state emission of neutrinos from point-sources typically require
a sample of neutrinos with $<10\%$ contamination by background atmospheric muons.
To achieve this goal we are limited to lower (Northern) hemisphere searches,
unless other time or energy-dependent constraints are included in the analysis (eg. for bursting emissions such as gamma-ray bursts (GRBs) or very high energy extra-galactic sources with much harder spectra than atmospheric neutrinos). The Earth acts as a filter and absorbs the background of atmospheric muons. It
begins to absorb neutrinos above the PeV range depending on the direction, which
determines the column density crossed by neutrinos.
The first results of IC9 are summarized in Sec.~\ref{sec1}, together with the 2005-6 data of AMANDA-II, which combined 
with the previous 5 years \cite{5year} constitutes the largest sample of high energy neutrinos ever collected for these studies
(924+887 in 386 d in 2005-6 + 4282 in 1001 d in 2000-4).
These results, together with the anticipated performance for various configurations of IceCube and expected and
measured results from other experiments are represented in Fig 1 (left), summarizing
the state of the art for neutrino point-like source searches.
\begin{figure}
\begin{tabular}{cc}
\includegraphics[width=7cm,height=7cm]{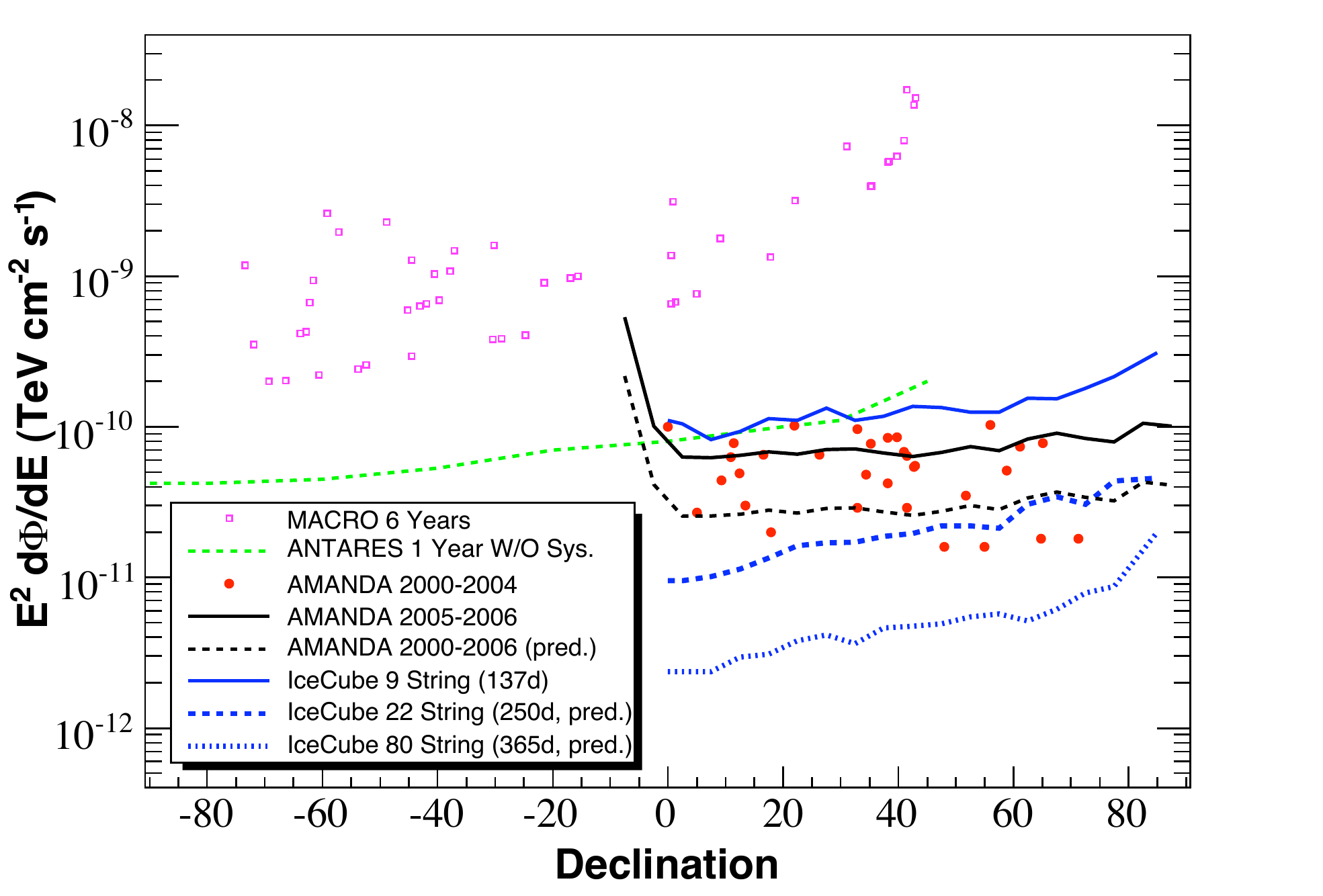}&
\includegraphics[width=7cm,height=7cm]{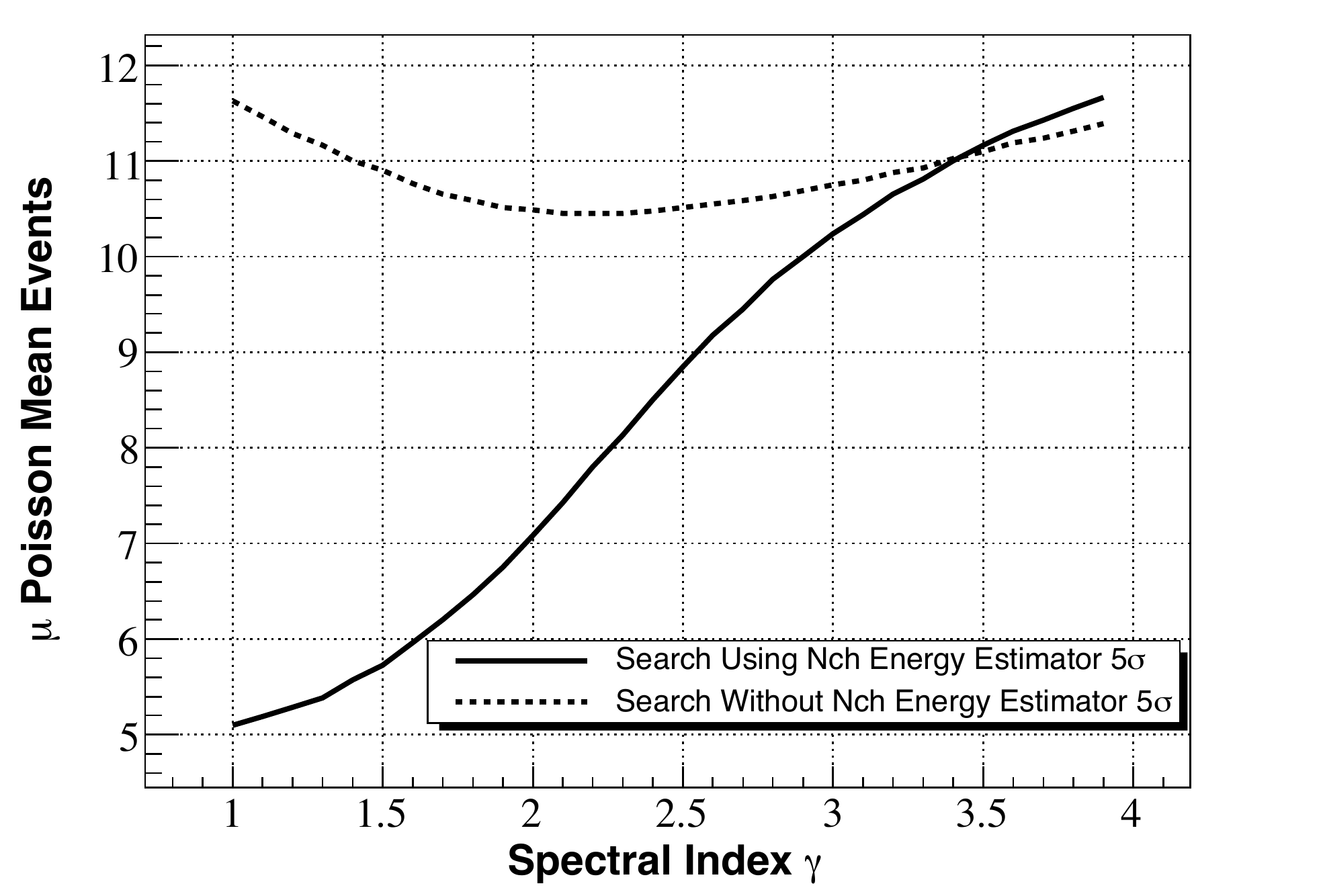}
\end{tabular}
\caption{\label{fig1} (left) Average (over right ascension) sensitivities vs declination for all sky point-source searches and upper limits for specific source catalogues. Neutrino fluxes are assumed to have $E^{-2}$ differential spectra. Results and predictions for Northern hemisphere experiments: MACRO upper limits \cite{macro} and predicted ANTARES sensitivity for 365 d \cite{antares}; South Pole experiments (from top to bottom): IC9 (137 d), AMANDA-II 2005-6 (386 d) analyses, dots are upper limits for AMANDA-II 2000-2004 \cite{5year} (1001 d). The upper dashed line is the sensitivity estimated
for the combined AMANDA-II data from 2000-6. IC22 is expected to take data for 250 d and the predicted sensitivity is better than the cumulative result of AMANDA-II (notice the 
strong trend with declination of IceCube compared to AMANDA-II due to the higher energy threshold that allows better discrimination of atmospheric neutrinos). The IC80 prediction for 365 d is also shown.
(right) Signal events needed for a neutrino source discovery with 50\% probability at 5$\sigma$ significance vs the spectral index of its differential flux using the Nch energy estimator (solid line) and not using it (dashed line) for a period equivalent to the 2005 data.}
\end{figure}

\section{IC9 and AMANDA-II results and expected performance of future configuration}
\label{sec1}
\begin{figure}
\begin{tabular}{cc}
\includegraphics[width=7cm,height=7cm]{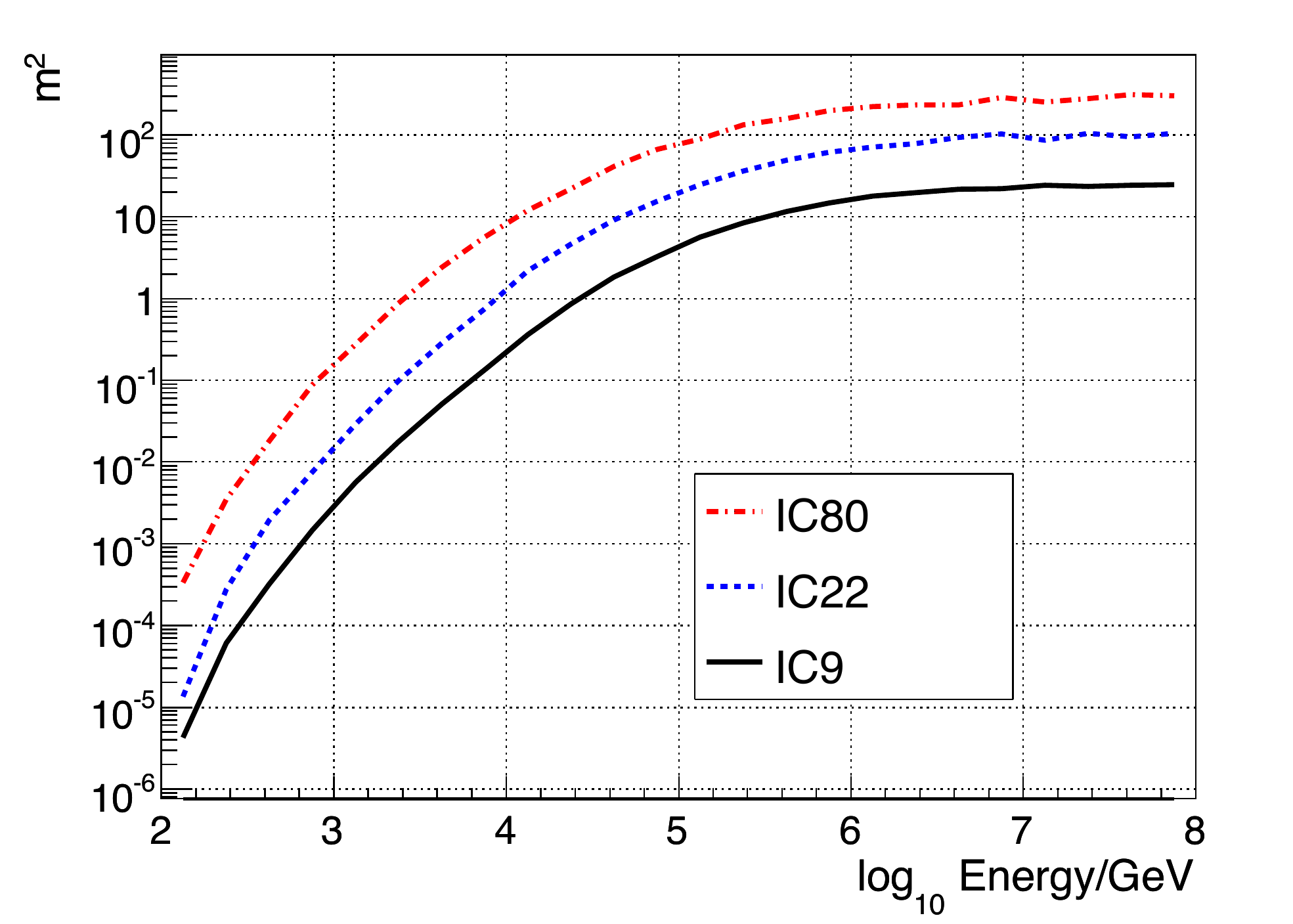}&
\includegraphics[width=7cm,height=7cm]{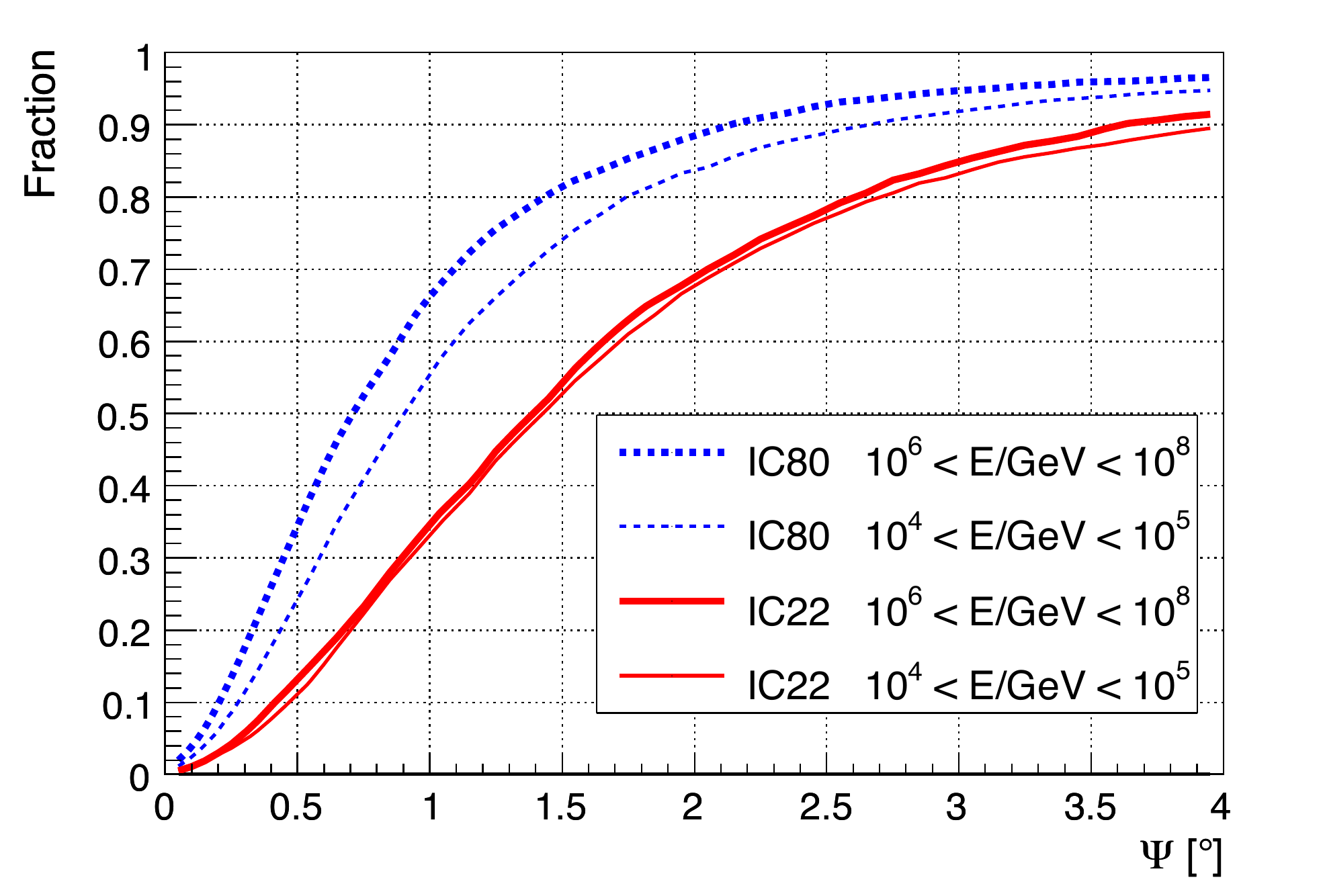}
\end{tabular}
\caption{\label{fig2} (left) Effective area averaged over the Northern hemisphere for 
equal contributions of neutrinos and antineutrinos as a function of energy for IC9, IC22 and IC80. (right) Point spread function (cumulative distribution of angular reconstruction errors) for IC22 and IC80 for two different energy ranges.
}
\end{figure}

The data sample for IC9 was obtained
in 137.4 d during Jun.-Nov. 2006: 233 neutrino events (227 expected from atmospheric neutrinos and less than 10\% from mis-reconstructed atmospheric muons) are extracted by a filtering procedure including two cuts aimed at rejecting coincident events from 2 independent showers (NDir = number of PMTs hit by unscattered light $> 8$  and $\sigma$ = event by event accuracy of tracking $<2.5^{\circ}$). More details on the analysis are in \cite{icrc2007}.
We have developed for these events and for the 2005-6 analysis of AMANDA-II a 
likelihood-based analysis that employs the angular distribution of the background + source hypothesis compared to the background only one. This method can be extended by including further information, as shown in Fig.~\ref{fig1}(right) where the two curves allow a comparison of the method using additional energy-based 
information and not using it. The variable used here that relates to the neutrino energy is Nch, the number of hit photomultipliers along the track. This variable is sensitive to different spectra and allows a better separation of signal from
background (atmospheric neutrinos follow an $E^{-3.6}$ power law above about 100 GeV) 
when the signal spectrum is harder. 
It can be seen that for an $E^{-2}$ source spectrum
the improvement is of the order of 30\% compared to the case using only the directional information, and the total improvement compared to more traditional binned analyses using
optimized angular windows \cite{5year} is about 45\% at declinations of $\sim 40^{\circ}$.
Similar studies will be performed using time dependent information especially in view of multi-wavelength studies involving the enormous amount of information coming from gamma-ray experiments. Other programs ongoing in the experiment are
Target of Opportunity programs in which IceCube sends alerts to other experiments \cite{icrc2007}.
In Fig~\ref{fig2} we show the effective area and point spread function for IC9 and what we expect for IC22 and IC80. 
These estimates are based on the selection criteria used in the IC9 analysis, while work continues on optimizing the cuts for the latest configuration.
With these cuts we expect about 3,700 events in IC22, and 40,000 in IC80 per year from atmospheric neutrinos from the lower hemisphere.

GRBs are extremely promising source candidates for 
IceCube, which will be at least 10 times more sensitive than AMANDA-II and should detect GRB neutrinos in a few years if fireball models are correct. We refer to \cite{icrc2007} for more details.

\section{Outlook}
Up to now there is no significant detection of astrophysical neutrinos.   
Nevertheless, we have well-developed methods to achieve a high discovery potential  
as the detector size approaches the cubic-kilometer scale.
At the moment the collaboration is considering  extending
the detector for high energy studies by deploying some of the outer strings
of the final detector at a larger distances than others. There is also interest to deploy an inner core array inside IceCube to enhance the detection capability for atmospheric neutrinos, WIMPs and galactic sources and lower the energy threshold down
to about 30 GeV.

\section*{References}

\title{Indirect Searches for Dark Matter with IceCube}
\author{Carsten Rott for the IceCube Collaboration}
\address{Pennsylvania State University, Department of Physics, 104 Davey Lab, University Park, PA 16802, USA}
\ead{carott@phys.psu.edu}
\begin{abstract}

The existence of dark matter can be inferred from a number of observations, among them
rotational profiles of galaxies, large scale structures, and WMAP's 
anisotropy measurement on the cosmic microwave background.
Weakly Interacting Massive Particles (WIMPs), `cold' thermal relics of the Big Bang, are 
leading dark matter candidates.
They are expected to be gravitationally trapped by massive bodies, such as the
Sun and the Earth, or objects at the Galactic Center, where they could then annihilate and produce
neutrinos, which could be detected by neutrino telescopes.
We describe searches performed in AMANDA for these events and the prospects for IceCube.

\end{abstract}

\section{Introduction}

In the concordance model of cosmology, only about $4\%$ of the universe's mass-energy content is
described by standard model particles.
The remaining fraction is predicted to be $23\%$ non-baryonic cold dark matter (DM) and $73\%$ dark energy.
Understanding the dark sector of the universe is one of the most challenging problems in modern physics.
Neutrino telescopes such as IceCube 
 can search indirectly for DM by detecting neutrinos created in DM annihilation processes.
IceCube will instrument a volume of $1~{\rm km}^3$ by 2011 with 80 strings, of which 22 have been deployed, each 
containing 60 Digital Optical Modules (DOMs).
The AMANDA detector, which has been taking data in its final configuration since 2000 
has been integrated into IceCube and is combinedly operated~\cite{ICRC_2007}.

\section{WIMP Searches with IceCube and AMANDA} 

Massive objects could gravitationally capture cold DM, resulting in an enhanced DM density at the center.
Neutrinos could then be produced in annihilation processes and would
escape from the source, providing a potential signal for kilometer scale neutrino telescopes.
Indirect searches for such DM signals are competitive to
direct searches for models with significant spin-dependent
neutralino-proton couplings~\cite{Halzen:2005ar}.

We have searched for such signals (from the Sun and center of the Earth) with AMANDA and upper limits were placed
on the neutrino flux produced by neutralino annihilation processes.

\begin{figure}[h]
\begin{minipage}{18pc}
\includegraphics[width=16pc]{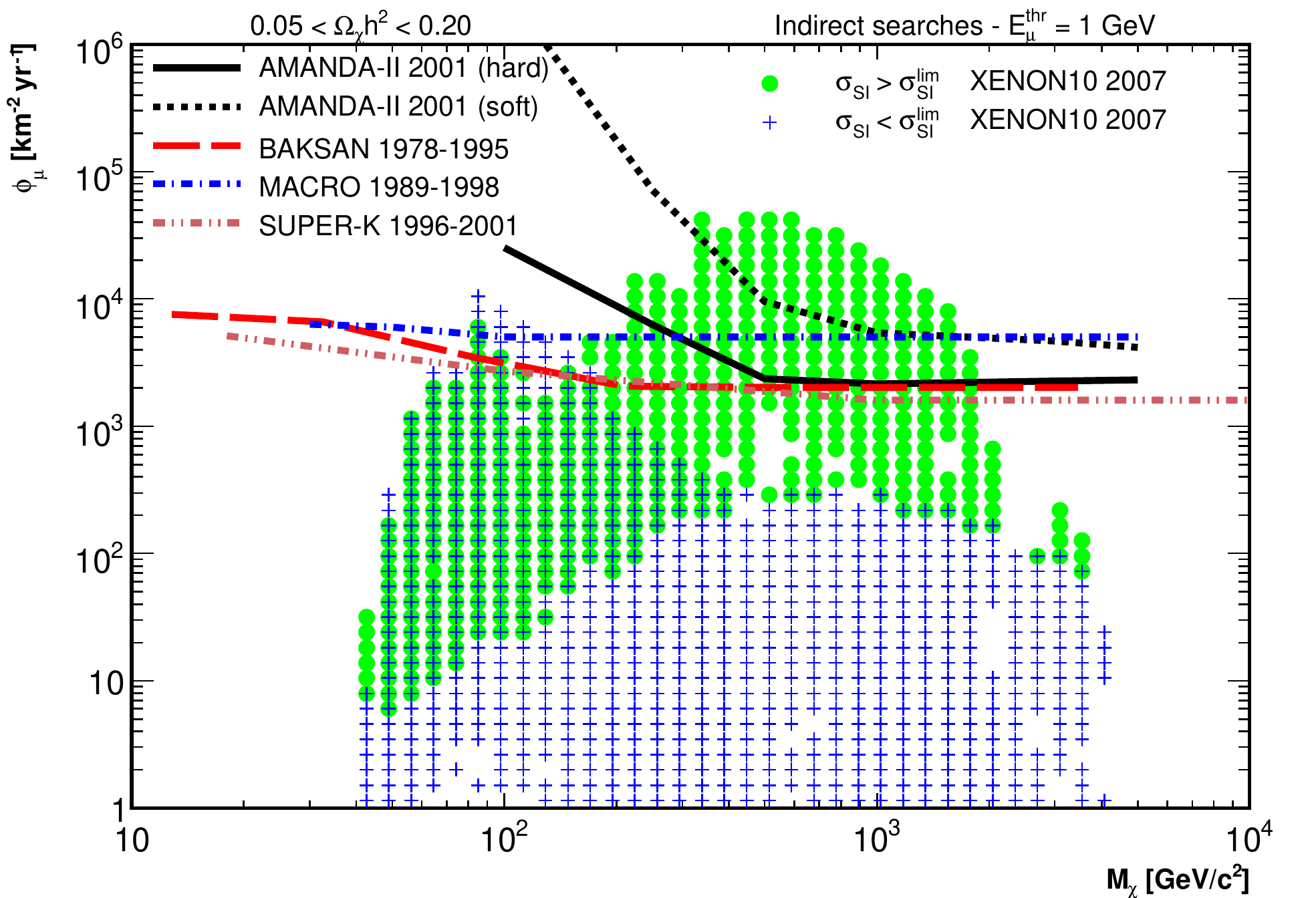}
\caption{\label{fig_sun}$90\%$~CL upper limit on the muon flux from neutralino annihilations in the Sun compared to other experiments.}
\end{minipage}\hspace{2pc}%
\begin{minipage}{18pc}
\includegraphics[width=16pc]{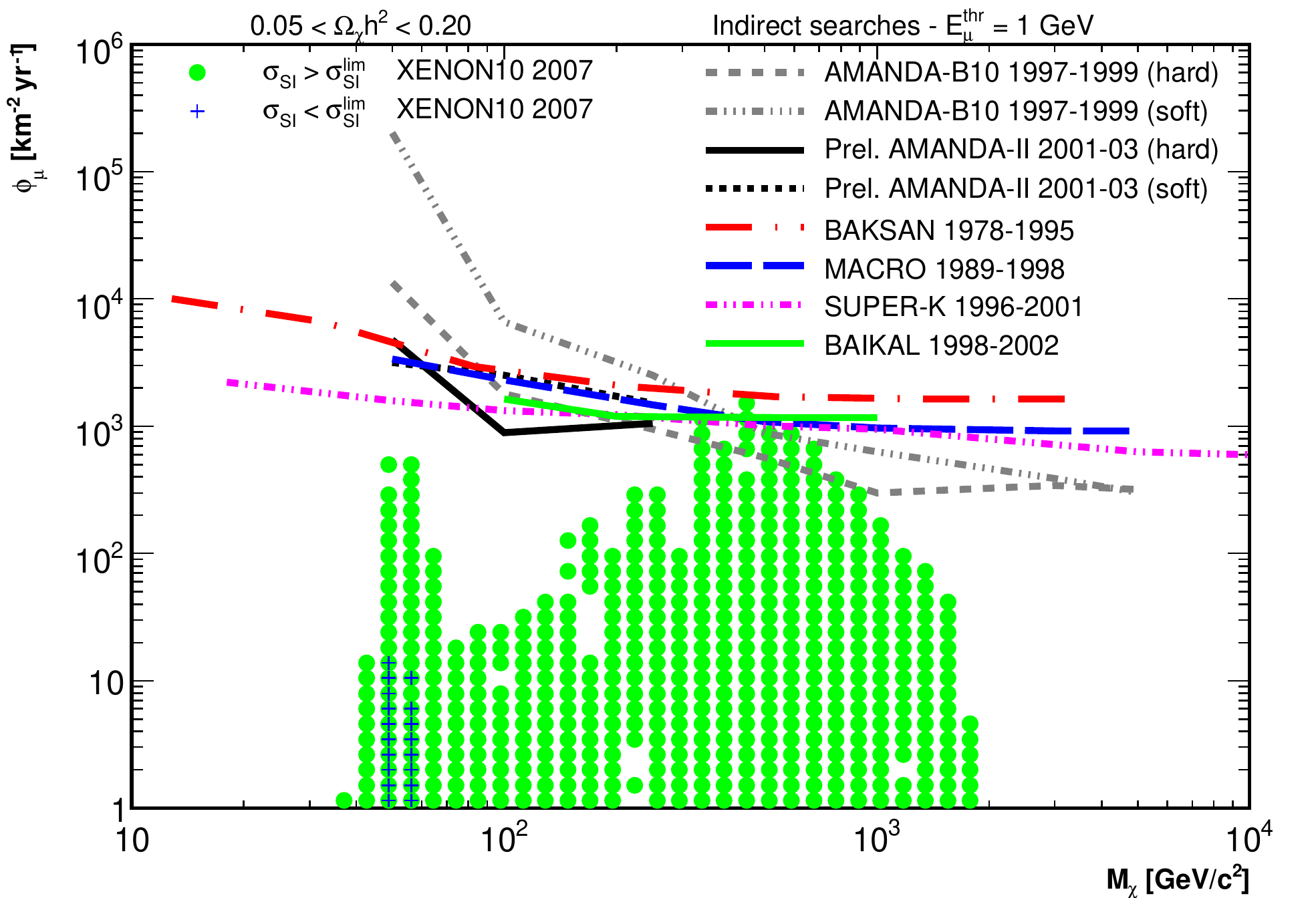}
\caption{\label{fig_earth}$90\%$~CL upper limit on the muon flux from neutralino annihilations in the Earth compared to other experiments.}
\end{minipage}
\end{figure}

In the search for solar WIMPs~\cite{solar_earth} the background was estimated from off-source data in the same declination band as the Sun. 
A $90\%$~CL exclusion 
limit on the neutrino flux was set for a neutralino mass range from $100~{\rm GeV}$ to $5~{\rm TeV}$ as seen in Fig.~\ref{fig_sun}.
Two annihilation channels were compared: ${\rm W}^+{\rm W}^-$, which produces a hard neutrino energy spectrum, and $b\bar{b}$ yielding a soft spectrum. 
The sensitivity for solar WIMPs has also been evaluated using the combined IceCube/AMANDA detector~\cite{ICRC_2007}
and a considerable improvement compared to AMANDA was achieved. 
The improvement is due to a reduced trigger threshold, a larger effective volume, and the
detector configuration, in which AMANDA forms a more densely instrumented subarray inside the IceCube detector, 
which is especially important for detection of sub-TeV neutrinos. 
Peripheral strings can be used as a veto region to more effectively reduce the atmospheric muon background. 
The detector configuration allows for an analysis of fully and partially contained events yielding additional sensitivity.

We have also searched for vertically up-going muon neutrinos as expected from neutralino annihilations 
in the Earth's core using the AMANDA dataset and placed a $90\%$~CL upper limit on the flux~\cite{solar_earth} as shown in Fig.~\ref{fig_earth}
and compared to other experiments.

The region around the Galactic Center is another good potential candidate for sources of neutrinos 
from DM annihilations. 
Located in the Southern sky, it was so far inaccessible to AMANDA due to an irreducible 
down-going atmospheric muon background.
However, the much larger IceCube detector might offer the potential to observe these neutrinos as fully or partially 
contained events. A dedicated filter for these contained events has been implemented~\cite{Rott_TeVIII}. 

\section{Conclusions and Outlook}

The combined IceCube/AMANDA detector provides a significant improvement in sensitivity for searches of DM. Dedicated
filters for these events have been implemented and the analysis effort is ongoing.
Extensions of IceCube to construct a more densly instrumented detector array near the bottom inside IceCube are also being studied.
Such a configuration would further improve the sensitifity to sub-TeV neutrinos, as the deep ice is clearer, the deeper location 
provides an improved shielding from cosmic ray backgrounds, and surrounding IceCube strings can be used 
to reject down-going atmospheric muons.

\title{Identification of Extremely-high energy starting neutrino events 
with the IceCube observatory}.


\author{M.~Ono, A.~Ishihara, and S.~Yoshida 
for the IceCube collaboration}


\address{Department of Physics, Chiba University, Chiba 263-8522, Japan}

\ead{mio@hepburn.s.chiba-u.ac.jp}

\begin{abstract}
The IceCube neutrino observatory is capable of
detecting Extremely High Energy (EHE) neutrinos with energies
beyond $10^8$ GeV originated in the highest energy cosmic rays.
The high energy muon bundles associated with cosmic ray air showers
constitute a major background and a reliable rejection mechanism must be
developed. A possible approach is to identify events 
induced by neutrinos inside the detector instrumentation volume.
The initial simulation study is briefly reported. 
\vspace{-1.5pc}
\end{abstract}

\section{Introduction}
Extremely high energy (EHE) neutrinos are expected to play a key role in
understanding the origin of EHE cosmic rays (EHECRs),
because energetic neutrinos can be produced by
collisions of EHECRs and the cosmic microwave background photons,
a process known as GZK mechanism~\cite{berezinsky69}.
In the initial stage, IceCube neutrino observatory~\cite{IceCubePhysics} has
performed the search for these EHE neutrinos by looking for extremely
luminous events as the signal~\cite{aya06, icrc07}.
Rejection and estimation of atmospheric muon background events is
central to the analysis. However, the simulation of atmospheric
muon events is associated with rather incomplete knowledges of the
hadronic interactions in the relevant energy range, which leads to a
hard-to-reduce uncertainty in muon background identification and
rejection.
One way to reduce this uncertainty is by selecting the neutrino events
which are induced inside the detector instrumented volume, which
in principle minimizes the dependence on our knowledge about the atmospheric muon bundles.

We report here the initial study of starting\footnotemark EHE neutrino event
identification with full IceCube using Monte Carlo simulation. 
The MC sample studied consists of isotropic muons and starting
$\nu_{\mu}$ events with energies from 10$^5$ GeV to 10$^{11}$ GeV, following
$E^{-1}$ spectrum. 
\footnotetext{Starting neutrino events are defined
as the ones with the neutrino interaction vertex point
within sphere of 400 m radius from the array center. }

\vspace{-0.5pc}
\section{The methods to identify neutrino-induced events}
\begin{figure}[t]
\vspace{-1pc}
   \begin{tabular}{cc}
    \begin{minipage}{8.5pc}
     \includegraphics*[width=4cm,height=4cm]{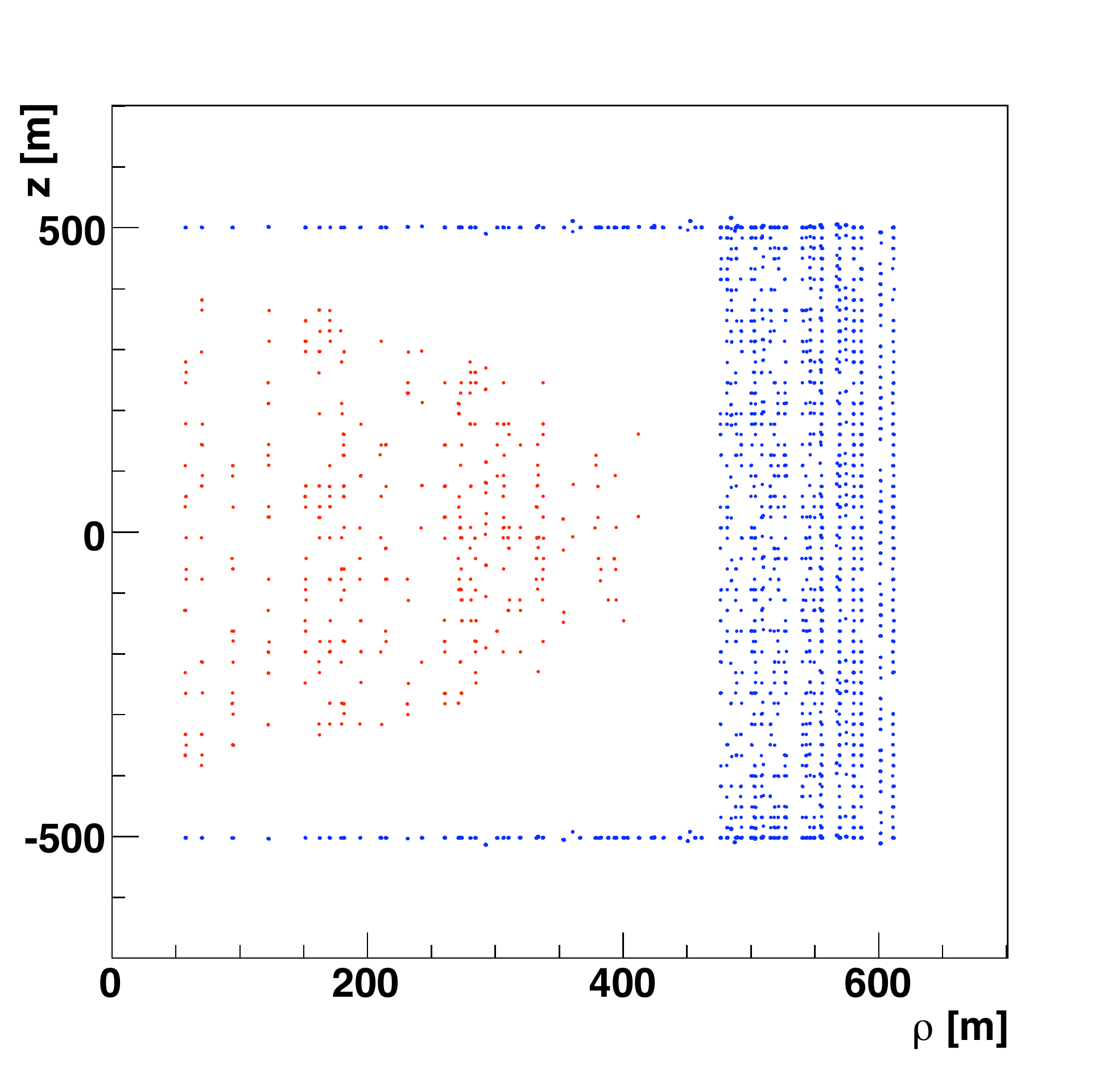}
     \noindent
     \vskip -1pc
     \caption{\label{fig_1stDom}Distribution of the first hit DOM from
     cylindrical coordinates. Blue dots are muon events and red dots are
     starting muon-neutrino events.}
      \end{minipage}
      \hspace{0.5pc}
      \begin{minipage}{29.pc}
       \begin{tabular}{ccc}
       \includegraphics*[width=3.5cm, height=4cm]{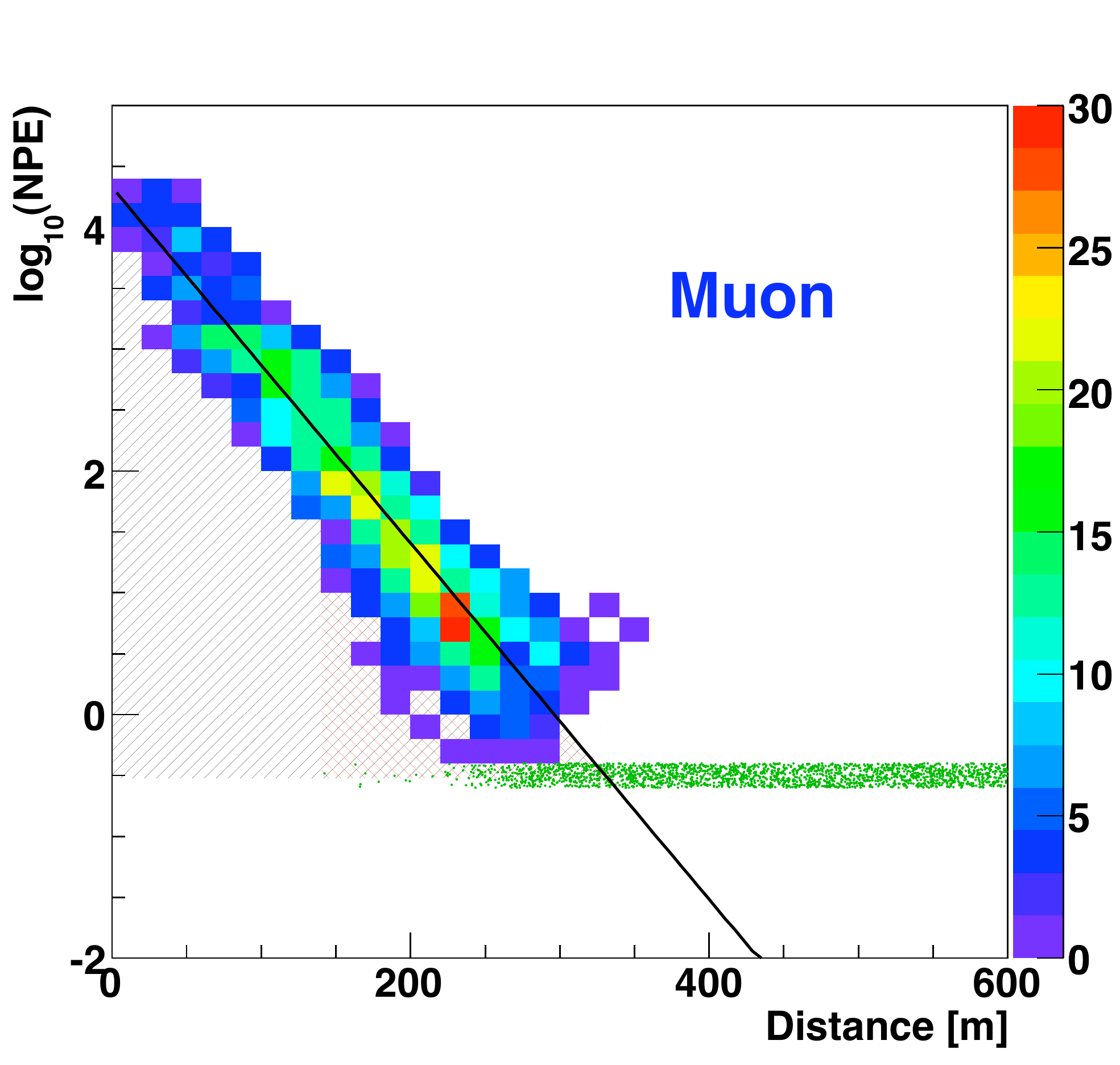} & \
       \includegraphics*[width=3.5cm, height=4cm]{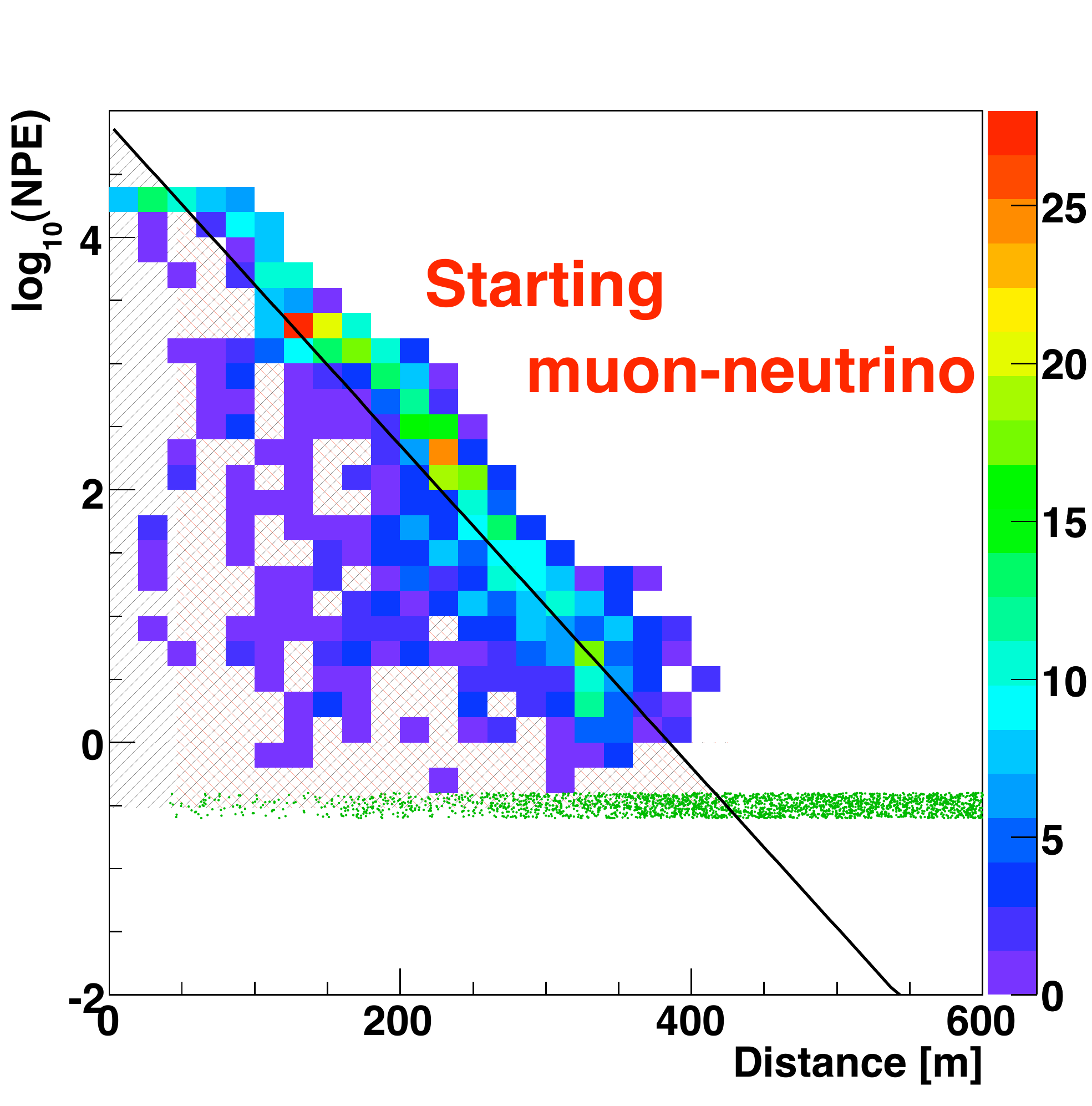}&\
       \includegraphics*[width=4cm, height=4cm]{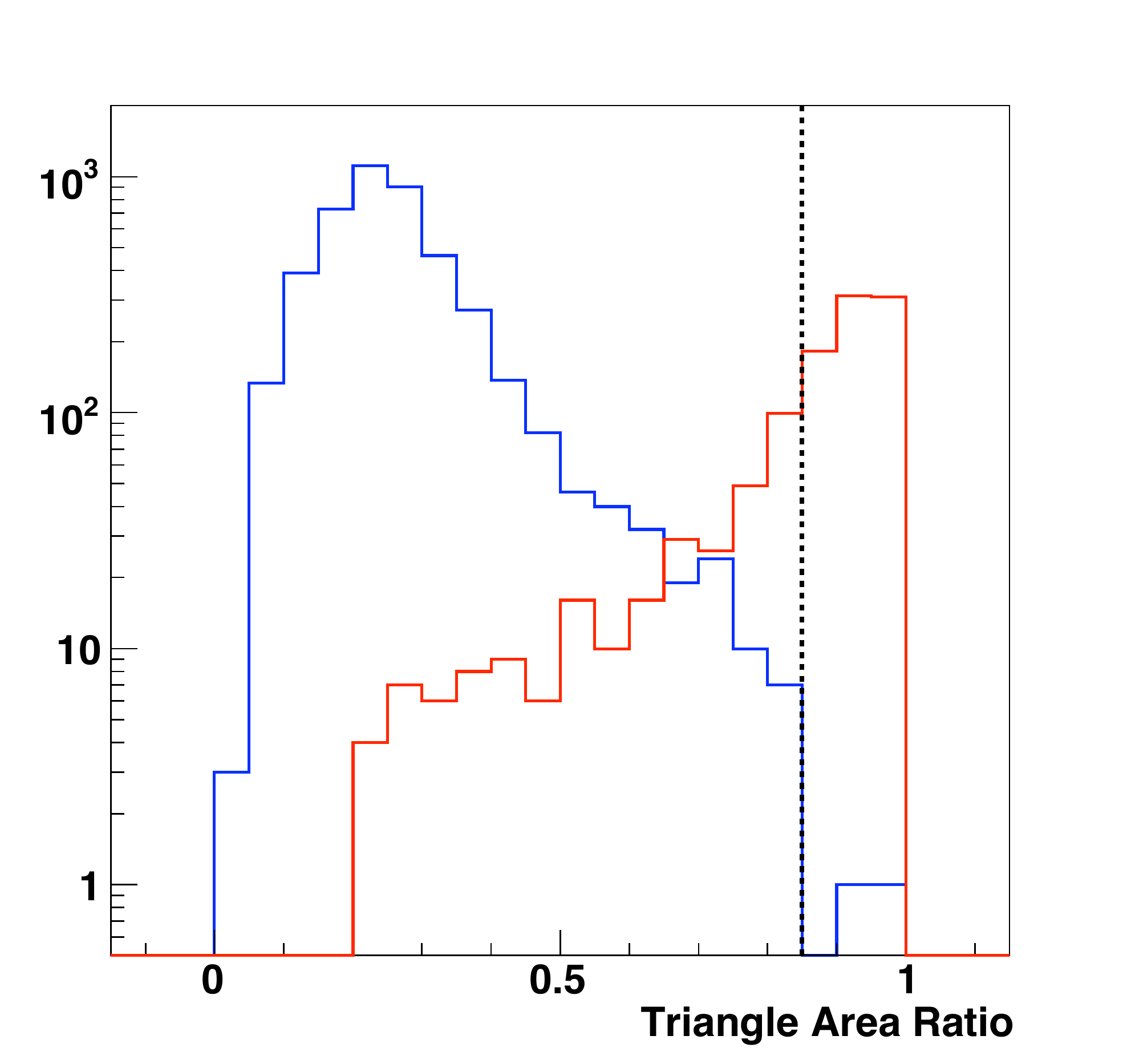}
       \end{tabular}
       \vskip -1pc
       \caption{\label{fig_lateral}Examples of the lateral distribution for a muon
       event (left) and a starting muon-neutrino event (middle). Green dots are
       DOMs bellow a threshold (0.3 photo-electron). Color histogram
       shows distribution of DOMs with signal. Black solid line is
       derived by fitting the distribution. The right triangles are
       shaded in black and red. Distribution of the triangle area ratio
       obtained by the lateral distribution is plotted in the right
       panel for though going muon events (blue) and the starting
       $\nu_{\mu}$ events (red). The triangle ratio cut is fixed at 0.85
       (black dotted line).}
      \end{minipage}
    \end{tabular}
\vspace{-1.1pc}
\end{figure}

We can expect that the location of $\nu_\mu$-N interaction is close to
the center of the IceCube detector, so are the positions of the IceCube
digital optical modules (DOMs) which first records the Cherenkov photons from the neutrino
induced charged lepton or hadron secondaries in the event, while the earliest hits in
atmospheric muon events are always found in the outer layer in-ice
detectors. 
%
%
Fig.~\ref{fig_1stDom} shows the distribution of
the first hit DOM locations of the simulated events
in cylindrical coordinates where $z$ is depth and $\rho$ the distance
from the central axis of the array.
Only the DOMs recording more than 5 photo-electrons
contribute in the plot to reduce the chance of mistaking the
earliest hit from noise as signal.
This threshold was determined by the real data
generated from nitrogen laser deployed in the deep ice
together with the DOMs~\cite{SC-ICRC}. One can see that
$\mu$ and starting $\nu_\mu$ events are clearly separated.

Another approach is based on the fact that, unlike muon
which emits Cherenkov photons throughout its track, 
trajectories of neutrinos are invisible to the optical sensors and thus
a starting neutrino induced event leaves sizable number of DOMs 
without photon signal near the neutrino trajectories. 
%
This can be observed in the characteristic lateral distributions.
Shown on the left and middle in Fig.~\ref{fig_lateral} is the number of
photo electrons recorded in each DOM as a function of lateral distance (LD) of the DOMs
from the MC-truth trajectories of the through-going $\mu$ and $\nu_{\mu}$
respectively.
The green band at the bottom of each distribution
represents those DOMs recording less than 0.3 photo-electron.
To describe the difference, we define the two right triangles 
shaded in black and red shown in the left panel of Fig.~\ref{fig_lateral}.
One has the right apex at
zero LD, and the other at the LD of the
closest LD DOMs with no photon signal recorded.
Taking the ratio of the areas of these triangles event-by-event,
it is observed that neutrino induced events are more likely to give this ratio
close to unity because DOMs without photon hit exist 
at smaller lateral distances. The area ratio distributions are shown in
the right panel of Fig.~\ref{fig_lateral}.
The plot indicates that this method identifies approximately 70 \%
of the starting neutrino events
with the area ration cut of 0.85, while rejecting 99.95 \% of muon events.

\vspace{-0.5pc}
\section{Discussion}
The first hit DOM method shows high efficiency but it relies
on a single DOM information and requires state-of-the-art level
of understanding of the detector response. The lateral
distribution method can be more reliable, but a high accuracy
of the track trajectory reconstruction is necessary. Note that
the present study uses the simulation truth trajectories.
We also found that irreducible background to this approach
is a muon decay event which very rarely occurs in EHE range.
A combination of the two methods described here might provide a more
reliable way to identify neutrino induced events
with less dependence on detailed detector response. 

\vspace{-0.5pc}
\section*{References}

\end{document}